\documentclass[11pt,a4paper]{article}
\usepackage[T1]{fontenc}
\usepackage{newunicodechar}
\newunicodechar{–}{\textendash}
\usepackage{jheppub}
\usepackage{mathtools}
\usepackage[protrusion=true,expansion=false]{microtype}
\newcommand{\dd}{\mathrm{d}}
\newcommand{\AdS}{\mathrm{AdS}}
\newcommand{\cF}{\mathcal{F}}
\newcommand{\sym}{\mathbin{\odot}}
\hypersetup{
  pdftitle={Emergence of Lorentzian geometry from string Carroll geometry},
  pdfauthor={Anirudhda Shinde and Mangesh Mandlik}
}

\title{Emergence of Lorentzian geometry from string Carroll geometry}
\author[a]{Anirudhda Shinde,}
\author[a]{Mangesh Mandlik}
\affiliation[a]{Department of Physics, Indian Institute of Technology (Indian School of Mines), Dhanbad, Jharkhand, India}
\emailAdd{24dr0031@iitism.ac.in}
\emailAdd{mandlik@iitism.ac.in}

\abstract{The near horizon region of a generic non extremal black hole has been extensively studied in terms of a Carrollian geometry with two null directions, also called a string Carroll geometry. We investigate how this string Carroll geometry evolves to a Lorentzian throat as the four dimensional Reissner–Nordström (RN) black hole approaches extremality. We identify the quadratic temporal contribution that is suppressed in the non extremal description but is required to recover the \(\mathrm{AdS}_2\) curvature. Approaching the outer horizon and extremality simultaneously, scaling the horizon separation together with the radial distance and correspondingly rescaling Killing time, brings this contribution into the leading longitudinal metric alongside the Rindler sector. In ingoing Eddington–Finkelstein coordinates, the string Carroll metric alone has a Lorentzian limit with a flat longitudinal sector. The quadratic temporal term arises from the bilinear of the leading and subleading longitudinal coframe fields and supplies the missing curvature, while higher radial contributions vanish. The reconstructed limiting geometry is already locally \(\mathrm{AdS}_2\times S^2\) at finite positive scaled horizon separation; setting this parameter to zero makes the selected horizon extremal without changing the local curvature. The limiting metric and Maxwell field satisfy the Einstein–Maxwell equations. In static coordinates, the temporal component again follows from the quadratic term, while the radial component requires summing contributions from all radial orders. The reconstruction provides a geometric starting point for investigating which expansion data must be retained when extending calculations based on the non extremal string Carroll description into the throat regime, including their implications for string, particle and field dynamics.}


\begin{document}
\maketitle

\section{Introduction}\label{sec:introduction}
The study of black holes often simplifies when one focuses on the region very close to the horizon. The full spacetime contains information about the black hole, its asymptotic region, and the coordinate system used to describe it.
Near the horizon, however, many of these details simplify, and one can isolate the local geometric structure that controls the physics seen by observers and probes close to the horizon. This is the reason why near horizon physics has become an important tool in the
study of black holes.

Near horizon limits are useful because they separate the local geometry near the horizon from the global structure of the spacetime. They help identify universal features shared by broad classes of black objects, even when the full solutions are different.
Such limits have played an important role in black hole thermodynamics \cite{Kunduri:2013gce,Jacobson:1995ab}, horizon symmetries \cite{Guica:2008mu,Compere:2012jk,Donnay:2015abr,Donnay:2016ejv,DeSimone:2025ouu,Chen:2010as}, quantum fields near horizons  \cite{Fredenhagen:1989kr}, and string-theoretic studies of black holes  \cite{Maldacena:1998uz,Strominger:1996sh}.

In many cases, the near horizon region is not just a technical approximation; it is the part of the geometry where the essential causal and physical features near the horizon become most transparent.
The nature of this near horizon region depends crucially on whether the black hole is extremal or non extremal. Extremal black holes have zero surface gravity and often admit a well-defined throat geometry in the near horizon limit \cite{Kunduri:2013gce}.
The standard example is the four dimensional extremal Reissner--Nordström (RN) black hole, whose near horizon geometry is \(AdS_2 \times S^2\) \cite{PhysRev.116.1331,Robinson:1959ev,Kunduri:2013gce}. 
This Lorentzian throat has its own geometric structure and has
been central in studies of black hole entropy, decoupling limits, and near horizon symmetries \cite{Sen:2005wa,Maldacena:1998uz,Kunduri:2013gce}. 
The corresponding near extremal RN black hole also has a
near horizon geometry described by a finite temperature \(AdS_2\) black hole patch, accompanied by the transverse \(S^2\) \cite{Spradlin:1999bn,Porfyriadis:2018jlw,Moitra:2018jqs}.

Non extremal black holes are different. Their horizons have nonzero surface gravity, and the immediate region outside the horizon is locally Rindler-like. This Rindler behaviour is universal for regular non extremal horizons \cite{Parikh:2012kg}. Unlike the extremal throat, however, the
non extremal near horizon region is usually best understood as an expansion around the horizon rather than as an independent throat geometry. The strict horizon limit is degenerate, and therefore one needs a careful way of carrying out the near horizon
expansion \cite{Fontanella:2022gyt}.

A natural language for such degenerate limits is Carrollian geometry. A Carrollian structure arises in an ultra-relativistic limit in which light cones close up and the metric becomes degenerate and loses its ordinary Lorentzian inverse structure \cite{Hansen:2021fxi,Ciambelli:2019lap,Duval:2014uoa,Hartong:2015xda,LevyLeblond:1965}.
This type of geometry appears naturally on null manifolds such as black hole horizons. In this sense, the horizon itself carries a Carrollian character: motion transverse to the null
direction becomes constrained, and the usual Lorentzian notion of propagation degenerates at the horizon \cite{Ciambelli:2019lap,Donnay:2019jiz}.
For near horizon regions rather than the horizon surface alone, the relevant structure is a little richer. In the non extremal case, the near horizon geometry contains two special longitudinal directions, forming the Rindler sector, together with transverse angular
directions such as the $S^{2}$ in four-dimensional black holes.
This leads to what is called a string Carroll geometry \cite{Bagchi:2023cfp,Bagchi:2024rje,Bagchi:2026qpi}. The terminology reflects the fact that, instead of a single Carrollian time direction, one has a two-dimensional longitudinal
sector and a transverse base.
For a spherically symmetric non extremal black hole, the
near horizon region can therefore be viewed as a string Carroll expansion in which a two-dimensional Rindler fibre is organised over the $S^{2}$.

This framework has been developed in recent work on strings and probes near non extremal black hole horizons \cite{Bagchi:2023cfp,Bagchi:2024rje,Bagchi:2026qpi}.
The central observation is that the small parameter
measuring the distance from the horizon plays the role of an effective Carrollian expansion
parameter. At leading order, the geometry becomes degenerate in the longitudinal directions, while the transverse sphere remains finite. First subleading order then contain corrections to this degenerate structure.
The analysis of non extremal near horizon regions in this string Carroll framework naturally raises a further question of how this structure behaves when an extremal limit is taken. The near horizon region of extremal black holes has already been studied
extensively and is known to have Lorentzian throat geometry rather than the Rindler-type structure of non extremal horizons \cite{Kunduri:2013gce,PhysRev.116.1331,Robinson:1959ev,Fontanella:2022gyt}. Therefore, once a non extremal near horizon region is
described in the string Carroll framework, it becomes important to understand how this near horizon string Carroll geometry evolves to a Lorentzian geometry on taking an extremal limit. This question was first identified in \cite{Bagchi:2026qpi,Bagchi:2024rje}. We address this question for RN by showing the emergence of the Lorentzian \(\mathrm{AdS}_2\) throat, starting from the string Carroll description of its non extremal near horizon region. While the string Carroll description of the non extremal region and the Lorentzian \(\mathrm{AdS}_2\) throat are both known, the present analysis establishes how these two geometric descriptions are connected as extremality is approached. Earlier string Carroll studies gave the familiar Rindler sector of non extremal horizons a geometric interpretation in terms of Carroll structures and developed a systematic framework for studying near horizon string dynamics. Our work pursues an analogous objective towards extremality, using RN as a concrete example.

In the string Carroll description of the non extremal region, the transverse
sphere remains finite, while the leading longitudinal contribution has Rindler
form. The quadratic temporal contribution is suppressed relative to the Rindler
term. Naively taking the horizon separation to zero does not recover the
\(\mathrm{AdS}_2\) throat. We instead approach the outer horizon and extremality
simultaneously, scaling the horizon separation together with the radial distance
from the outer horizon\footnote{In what follows, we refer to this scaling of the
horizon separation as the near extremal scaling.} and correspondingly rescaling
Killing time. This correlated scaling brings the Rindler term and the curvature restoring quadratic
term into the leading longitudinal metric. The accompanying correction to the
quadratic coefficient, proportional to the horizon separation, remains subleading
and vanishes in the limit. Related subtleties of extremal limits have appeared
in other contexts as well \cite{Carroll:2009maa,Bengtsson:2014fha}.

In ingoing Eddington–Finkelstein coordinates, applying the correlated limit described above to the string Carroll metric gives a nondegenerate Lorentzian geometry with a flat longitudinal sector. Retaining the quadratic temporal contribution restores the \(\mathrm{AdS}_2\) curvature. We trace this correction to the bilinear of the leading and subleading longitudinal fields in the RN coframe expansion. Higher radial orders vanish under the same correlated limit, together with their derivatives in the rescaled coordinates. The second order near horizon expansion is therefore sufficient to recover the curved throat.

The reconstructed geometry is locally $\mathrm{AdS}_2\times S^2$ for every finite scaled horizon separation $a\geq0$. For $a>0$, it describes the near extremal RN throat, whose horizon has nonzero surface gravity. Setting $a=0$ gives the extremal throat without changing the local curvature. Thus AdS curvature is already present in the near extremal throat, and the family of throat metrics extends smoothly to its extremal member in ingoing coordinates.

The reconstruction also makes precise the regularity of the limiting procedure. In the rescaled ingoing coordinates, the metric and its derivatives approach the throat smoothly on local coordinate regions, including at the horizon. A regular limiting coframe is obtained after a compensating Lorentz boost. The time rescaling and the boost themselves become singular at the endpoint, so this regularity concerns the reconstructed geometry and does not imply separate regular limits for the original string Carroll coframes. The relation between the two descriptions is thus established by following the expansion through the correlated scaling.

We also follow the electromagnetic field through the same procedure. For the chosen Maxwell potential, regular at the outer horizon, the first order radial expansion suffices to recover the throat electric field. The reconstructed metric and electromagnetic field satisfy the Einstein–Maxwell equations, whereas the limiting first order metric fails their trace equation. This provides an independent test of the temporal correction and shows that the metric and the regular Maxwell potential require different orders of the non extremal expansion.

We also examine how the throat emerges in static coordinates. The quadratic temporal correction again recovers the time component of the throat metric, but the radial component cannot be obtained from any finite truncation of the non extremal near horizon expansion. In this sector, the coefficients contain inverse powers of the horizon separation, allowing contributions from every radial order to survive the correlated limit. Recovering the radial component therefore requires summing these contributions. This gives the static form of the same Lorentzian throat obtained in the Eddington–Finkelstein analysis. The comparison shows that the reconstruction from the non extremal expansion proceeds differently in the two coordinate systems, even though the resulting geometry is the same.

This connection provides a concrete way to examine the adequacy of calculations based on the non extremal string Carroll description as extremality is approached. The RN analysis shows that a contribution suppressed in this description determines the longitudinal curvature of the leading limiting geometry and can consequently affect the leading equations and constraints of probes. The reconstruction therefore provides a geometric starting point for extending calculations based on non extremal near horizon expansions into the throat regime and investigating which expansion data are required for the physical quantity or dynamical theory under consideration. For string, particle and field dynamics, a complete treatment requires following the dynamical variables through the same scaling as the background geometry.

The paper is organized as follows. Section~\ref{sec:setup} introduces
the geometric expansion, identifies the RN near horizon data and
specifies the correlated scaling. Section~\ref{sec:curvature}
reconstructs the Lorentzian throat in ingoing coordinates and examines
the corresponding coframe limit. Section~\ref{sec:maxwell} treats the
Maxwell field and verifies the limiting Einstein--Maxwell equations.
Section~\ref{sec:static} studies the static expansion and the
summation of contributions from all radial orders. Section~\ref{sec:discussion}
summarizes the results and discusses the implications for probe dynamics
and further applications.
Appendix~\ref{app:curvature} gives the longitudinal curvature
calculation, while Appendix~\ref{app:sc-symmetries} records the
transformations of the expanded background fields.

\section{The non extremal near horizon expansion and the correlated limit}\label{sec:setup}

\subsection{Geometric structure of the string Carroll expansion}\label{sec:sc-geometry}

The string Carroll expansion distinguishes a two dimensional
longitudinal sector from the transverse geometry. In a frame
adapted to a non extremal black hole horizon, the longitudinal
sector describes the time and radial directions, while the
transverse sector describes the spatial horizon cross section.
Now to identify which string Carroll fields supply the contribution
needed to recover the extremal RN throat curvature, we express the metric
through $c^4$ in terms of the coframes and their bilinears.
We use the vielbein formulation
presented in ~\cite{Bagchi:2024rje,Banerjee:2025bkg}.

We work locally in four dimensions with signature $(-,+,+,+)$. Let $T^A=T^A_\mu\dd x^\mu$ and $E^i=E^i_\mu\dd x^\mu$ be longitudinal and transverse one forms, with $A,B=0,1$, $i,j=2,3$, and internal metrics $\eta_{AB}=\operatorname{diag}(-1,1)$ and $\delta_{ij}$. We take the coframe $(T^A,E^i)$ to be regular and
invertible at $c=0$. Its dual frame, obtained by coframe inversion, has components $U_A^\mu$ and $W_i^\mu$. For $c>0$, the physical Lorentz coframe is $(cT^A,E^i)$. We define the longitudinal and transverse bilinears and write
the corresponding expressions for the metric and its inverse:
\begin{equation}
 \begin{aligned}
 \mathcal L_{\mu\nu}&=\eta_{AB}T^A_\mu T^B_\nu,&
 \Pi_{\mu\nu}&=\delta_{ij}E^i_\mu E^j_\nu,\\
 \mathcal L^{\mu\nu}&=\eta^{AB}U_A^\mu U_B^\nu,&
 \Pi^{\mu\nu}&=\delta^{ij}W_i^\mu W_j^\nu,\\
 g_{\mu\nu}(c)&=c^2\mathcal L_{\mu\nu}+\Pi_{\mu\nu},&
 g^{\mu\nu}(c)&=c^{-2}\mathcal L^{\mu\nu}+\Pi^{\mu\nu}.
 \end{aligned}
 \label{eq:sc-split}
\end{equation}
By construction, the coframe and its dual satisfy
\begin{equation}
 \begin{split}
 T^A_\mu U_B^\mu&=\delta^A{}_B,\qquad
 E^i_\mu W_j^\mu=\delta^i{}_j,\qquad
 T^A_\mu W_i^\mu=E^i_\mu U_A^\mu=0,\\
 T^A_\mu U_A^\nu+E^i_\mu W_i^\nu&=\delta_\mu{}^\nu.
 \end{split}
 \label{eq:sc-complementarity}
\end{equation}
Thus $\mathcal L$ and $\Pi$ describe complementary longitudinal and transverse subspaces. The parameter $c$ is the conventional parameter of the Carroll contraction. Its relation to radial proximity will be identified in the RN expansion below.
We expand the coframe and its dual in powers of $c^2$
\begin{equation}
 \begin{split}
 T^A&=\tau^A+c^2m^A+O(c^4),\\
 E^i&=e^i+c^2e_{(1)}^i+c^4e_{(2)}^i+O(c^6),\\
 U_A^\mu&=u_A^\mu+c^2u_{(1)A}^\mu+O(c^4),\qquad
 W_i^\mu=w_i^\mu+O(c^2).
 \end{split}
 \label{eq:sc-frame-fields}
\end{equation}
Parenthesized subscripts label expansion coefficients. The dual coefficients follow by expanding~\eqref{eq:sc-complementarity}.\\
Substitution into the decomposition~\eqref{eq:sc-split} gives
\begin{equation}
 \begin{split}
 g_{\mu\nu}(c)&=h_{\mu\nu}+c^2(k_1)_{\mu\nu}
                         +c^4(k_2)_{\mu\nu}+O(c^6),\\
 g^{\mu\nu}(c)&=c^{-2}u^{\mu\nu}+\bar h^{\mu\nu}+O(c^2).
 \end{split}
 \label{eq:sc-expansion}
\end{equation}
The covariant coefficients are bilinears of the same coframe fields. With $\alpha\sym\beta=(\alpha\otimes\beta+\beta\otimes\alpha)/2$, they are
\begin{equation}
 \begin{split}
 h&=\delta_{ij}e^i\otimes e^j,\qquad
 \ell=\eta_{AB}\tau^A\otimes\tau^B,\\
 \chi&=2\delta_{ij}e^i\sym e_{(1)}^j,\qquad k_1=\ell+\chi,\\
 k_2&=2\eta_{AB}\tau^A\sym m^B
 +\delta_{ij}\bigl(e_{(1)}^i\otimes e_{(1)}^j
                      +2e^i\sym e_{(2)}^j\bigr).
 \end{split}
 \label{eq:sc-k2}
\end{equation}
The expression for $k_2$ identifies the contribution of the
subleading longitudinal field $m^A$ to the metric through
the bilinear $2\eta_{AB}\tau^A\odot m^B$. In the RN analysis
below, we show that the correlated limit of this contribution
supplies the quadratic temporal term responsible for the
$\mathrm{AdS}_2$ curvature of the throat.\\
Similarly, the bilinears of the dual frame coefficients
give\footnote{Parentheses on spacetime indices denote symmetrization with weight one half.}
\begin{equation}
 \begin{split}
 u^{\mu\nu}&=\eta^{AB}u_A^\mu u_B^\nu,\qquad
 h_\perp^{\mu\nu}=\delta^{ij}w_i^\mu w_j^\nu,\\
 \bar h^{\mu\nu}&=h_\perp^{\mu\nu}
            +2\eta^{AB}u_A^{(\mu}u_{(1)B}^{\nu)}.
 \end{split}
 \label{eq:sc-inverse-fields}
\end{equation}
The leading part of~\eqref{eq:sc-complementarity} implies
\begin{equation}
 \begin{split}
 h_{\mu\rho}u^{\rho\nu}&=0,\qquad
 \ell_{\mu\rho}h_\perp^{\rho\nu}=0,\\
 \ell_{\mu\rho}u^{\rho\nu}
 +h_{\mu\rho}h_\perp^{\rho\nu}&=\delta_\mu{}^\nu.
 \end{split}
 \label{eq:sc-leading-relations}
\end{equation}
The pair $(h_{\mu\nu},u^{\mu\nu})$ defines the leading string Carroll geometry~\cite{Bagchi:2024rje}. The tensor $h$ has signature $(0,0,+,+)$ and is degenerate along the longitudinal directions; $u$ retains their Lorentzian structure with signature $(-,+,0,0)$. 

Expanding the local frame transformations determines the
transformation laws of the string Carroll fields. These
transformations leave the leading pair $(h,u)$ and the full
metric coefficients $k_1$ and $k_2$ unchanged, although their
separate longitudinal and transverse contributions can
transform. Appendix~\ref{app:sc-symmetries} records these laws. In the RN application, we use the same fields to identify the contribution of $k_2$ that survives the correlated throat limit.

\subsection{The RN geometry and its near horizon data}\label{sec:rn-carroll}

We use geometrized RN parameters $M$ and $Q$, with $M>|Q|$ for a non extremal seed. The horizon radii and surface gravity are
\begin{equation}
 r_\pm=M\pm\sqrt{M^2-Q^2},\qquad
 \delta=r_+-r_->0,\qquad \kappa_+=\frac{\delta}{2r_+^2}.
 \label{eq:horizon-data}
\end{equation}
In ingoing coordinates, with $x=r-r_+$, the RN metric is~\cite{Reissner1916,Nordstrom1918,townsend1997blackholes}
\begin{equation}
 \dd s^2=-f(x)\dd v^2+2\dd v\dd x+(r_++x)^2\dd\Omega_2^2,
 \qquad f(x)=\frac{x(x+\delta)}{(r_++x)^2},
 \label{eq:seed}
\end{equation}
where $\dd\Omega_2^2=\dd\theta^2+\sin^2\theta\dd\phi^2$. The coordinate $x$ is an affine radial coordinate in this chart, rather than a proper spatial distance. The Killing generator is $\partial_v$. Pulling the metric back to the horizon $\mathcal H=\{x=0\}$ gives
\begin{equation}
 \left.\dd s^2\right|_{\mathcal H}=r_+^2\dd\Omega_2^2.
 \label{eq:rn-horizon-pullback}
\end{equation}
On the three dimensional horizon this induced metric is degenerate along $\partial_v$. The near horizon expansion instead retains the radial direction and yields the string Carroll structure on a four dimensional coordinate domain~\cite{Bagchi:2026qpi}.

To describe this neighbourhood, write $x=\epsilon R$, where $\epsilon>0$ is dimensionless and $\epsilon\to0$ at fixed $R$ approaches the outer horizon. With $r_+$ and $\delta>0$ fixed and $v$ unchanged, we identify
$\epsilon=c^2$ and write the expansion in the notation
of~\eqref{eq:sc-expansion}. The coefficients $h$, $k_1$, and
$k_2$ then take their RN forms:\footnote{The subscript $\epsilon$ makes the dependence of the metric on the
near horizon expansion parameter explicit.}
\begin{align}
 \dd s_\epsilon^2&=h+\epsilon k_1+\epsilon^2 k_2+O(\epsilon^3),\label{eq:expansion}\\
 h&=r_+^2\dd\Omega_2^2,\notag\\
 k_1&=2\dd v\dd R-\frac{\delta R}{r_+^2}\dd v^2+2r_+R\dd\Omega_2^2,\notag\\
 k_2&=-R^2\left(\frac{1}{r_+^2}-\frac{2\delta}{r_+^3}\right)\dd v^2
       +R^2\dd\Omega_2^2.\notag
\end{align}
We refer to~\eqref{eq:expansion} as the non extremal
near horizon expansion and use the term string Carroll
metric for its first order part $h+\epsilon k_1$.
For the split into the longitudinal $(v,R)$ sector and the
transverse sphere, the bilinears $\ell$ and $\chi$
in~\eqref{eq:sc-k2} are
\begin{equation}
 \begin{split}
 \ell&=2\dd v\dd R-2\kappa_+R\dd v^2,\\
 \chi&=2r_+R\dd\Omega_2^2.
 \end{split}
 \label{eq:rn-sc-identification}
\end{equation}
The bilinear $\ell$ is the two dimensional Rindler metric associated with the non extremal horizon, while $\chi$ is the first correction to the transverse sphere. Thus the near horizon parameter $\epsilon$ plays the role
of the Carroll expansion parameter $c^2$: the transverse
metric remains at leading order, while the longitudinal
sector first contributes at order
$\epsilon$~\cite{Bagchi:2023cfp,Bagchi:2024rje}.\\ 
The complementary contravariant data have nonzero components
\begin{equation}
 u^{vR}=u^{Rv}=1,\qquad u^{RR}=2\kappa_+R,\qquad
 h_\perp^{\theta\theta}=\frac{1}{r_+^2},\qquad
 h_\perp^{\phi\phi}=\frac{1}{r_+^2\sin^2\theta}.
 \label{eq:rn-sc-dual}
\end{equation}
They satisfy~\eqref{eq:sc-leading-relations} directly. In particular, $h$ is degenerate along both $\partial_v$ and $\partial_R$ on the four dimensional near horizon coordinate domain, while $u^{\mu\nu}$ retains the longitudinal causal structure.

The RN bilinears above can also be obtained by expanding
a coframe for the RN metric. Expressing this coframe in
the form~\eqref{eq:sc-frame-fields} makes explicit which
longitudinal and transverse fields contribute to the
second order coefficient $k_2$. On an angular patch, choose one forms $\bar e^i$ satisfying $\delta_{ij}\bar e^i\otimes\bar e^j=\dd\Omega_2^2$. In a null longitudinal basis, with
$\eta_{+-}=\eta_{-+}=1$ and $\eta_{++}=\eta_{--}=0$,
we choose the coframe
\begin{equation}
 T^+=\dd v,\qquad
 T^-=\dd R-\frac{f(\epsilon R)}{2\epsilon}\dd v,\qquad
 E^i=(r_++\epsilon R)\bar e^i.
 \label{eq:rn-frame-exact}
\end{equation}
It gives $g_\epsilon=2\epsilon T^+\sym T^-+\delta_{ij}E^i\otimes E^j$. Expanding at fixed seed parameters and using $\epsilon=c^2$ yields
\begin{equation}
 \begin{split}
 \tau^+&=\dd v,\qquad
 \tau^-=\dd R-\frac{\delta R}{2r_+^2}\dd v,\qquad m^+=0,\\
 m^-&=-\frac12\left(\frac1{r_+^2}-\frac{2\delta}{r_+^3}\right)R^2\dd v,\\
 e^i&=r_+\bar e^i,\qquad e_{(1)}^i=R\bar e^i,\qquad e_{(2)}^i=0.
 \end{split}
 \label{eq:rn-frame-coefficients}
\end{equation}
Consequently, the temporal part of $k_2$ is
$2\tau^+\sym m^-$, while its angular part comes from
$\delta_{ij}e_{(1)}^i\otimes e_{(1)}^j$. 
Together these express $k_2$ (in~\eqref{eq:expansion}) entirely through bilinears of the coframe fields, showing that the correction beyond the string Carroll metric arises from the same underlying vielbein expansion.
In the string Carroll expansion, $\epsilon^2 k_2$ is a
suppressed higher order contribution contained in the
$O(\epsilon^2)$ remainder. We show below that, under the
combined near extremal scaling and time rescaling, its
temporal part contributes at leading order in the resulting
limit and supplies the $\mathrm{AdS}_2$ curvature of the
throat.

\subsection{A family approaching extremality}

A naive extremal limit of the string Carroll metric, obtained by sending $\delta\to0$, removes the $\dd v^2$ term without producing the quadratic radial dependence of the RN throat. To investigate how the throat emerges from the non extremal string Carroll description, we approach the outer horizon and extremality simultaneously, scaling the horizon separation together with the radial distance and rescaling Killing time:
\begin{equation}
 \delta=\epsilon a,\qquad
 r_\pm=r_0\pm\frac{\epsilon a}{2}, \qquad v=\frac{V}{\epsilon},\qquad
 \epsilon\longrightarrow0,
 \label{eq:scaling}
\end{equation}
at fixed $r_0>0$, finite $a>0$, and fixed $V,R,\theta,\phi$.
We call $\delta=\epsilon a$ the near extremal scaling and
refer separately to $v=V/\epsilon$ as the time rescaling.
Setting $a=0$ after taking the limit gives the extremal
member of the resulting throat family.

\section{The contribution that restores AdS curvature}\label{sec:curvature}

\subsection{What the first order metric retains}

Applying ~\eqref{eq:scaling} to $\dd s_{[1],\epsilon}^2=h+\epsilon k_1$, its limiting metric becomes\footnote{Square brackets indicate the order retained in the
non extremal expansion, and this label is preserved after taking
the limit. The subscript $a$ labels the resulting family of
limiting metrics.}
\begin{equation}
 \dd s_{[1],a}^2=2\dd V\dd R-\frac{aR}{r_0^2}\dd V^2+r_0^2\dd\Omega_2^2.
 \label{eq:first-limit}
\end{equation}
For a longitudinal metric $-F(R)\dd V^2+2\dd V\dd R$, the determinant is $-1$ and its scalar curvature is $\mathcal R^{(2)}=-F''(R)$; a derivation is given in appendix~\ref{app:curvature}. Thus~\eqref{eq:first-limit} is already Lorentzian and nondegenerate, but
\begin{equation}
 \mathcal R^{(2)}_{[1]}=0.
 \label{eq:first-curvature}
\end{equation}
For $a>0$, its longitudinal factor is a Rindler region of flat spacetime. At $a=0$ it remains flat. The limiting operation exists at first order, but does not produce the curved RN throat.\\
The missing term is apparent in the numerator $x(x+\delta)$
of the blackening factor $f(x)$. In the string Carroll description of the non extremal RN near horizon
region, the radial coordinate is scaled as $x=\epsilon R$, but the horizon separation $\delta$ is held fixed. The term
$\delta x$ in the numerator supplies the temporal part
of the Rindler sector, which is subleading relative to the transverse
geometry. The quadratic term $x^2$ is further suppressed and lies
beyond the first order string Carroll metric. However under ~\eqref{eq:scaling}, we obtain
\begin{equation}
\delta x=\epsilon^2 aR,\qquad
x^2=\epsilon^2R^2,\qquad
\dd v^2=\epsilon^{-2}\dd V^2.
\label{eq:competition}
\end{equation}
Both temporal contributions therefore remain finite and enter the
leading longitudinal metric of the throat, with the quadratic term
restoring the $\mathrm{AdS}_2$ curvature.

\subsection{Second order completion and the Lorentzian throat}\label{sec:completion}

The second order contribution in~\eqref{eq:expansion} has the limit
\begin{equation}
 \epsilon^2 k_2\longrightarrow-\frac{R^2}{r_0^2}\dd V^2.
 \label{eq:second-survives}
\end{equation}
By~\eqref{eq:rn-frame-coefficients}, this is the limit of the longitudinal contribution $2\epsilon^2\tau^+\sym m^-$ in the chosen RN coframe. Its part proportional to $\delta$, as well as the transverse contribution $\epsilon^2R^2\dd\Omega_2^2$, vanishes after~\eqref{eq:scaling}. Combining~\eqref{eq:first-limit} and~\eqref{eq:second-survives} gives
\begin{equation}
 \boxed{\dd s_a^2=2\dd V\dd R-F_a(R)\dd V^2+r_0^2\dd\Omega_2^2,
 \qquad F_a(R)=\frac{R(R+a)}{r_0^2}.}
 \label{eq:throat}
\end{equation}
The longitudinal determinant is again $-1$, while
\begin{equation}
 \mathcal R^{(2)}_a=-\frac{2}{r_0^2}.
 \label{eq:throat-curvature}
\end{equation}
The metric is therefore locally $\AdS_2(r_0)\times S^2(r_0)$ for every finite $a\geq0$, with the $a>0$ members describing the near extremal RN throat~\cite{Spradlin:1999bn,Sen:2011cn}. The horizon $R=0$, generated by $\partial_V$, has surface
gravity $\widehat\kappa=F_a'(0)/2=a/(2r_0^2)$.
Setting $a=0$ makes this horizon extremal. Thus the quadratic contribution restores the AdS curvature already at finite scaled horizon separation.

For the RN function $f(x)$, the remainder beyond second
radial order is $O(x^3)$, uniformly as $\delta\to0$ and
$r_+\to r_0>0$. Under radial scaling and ~\eqref{eq:scaling}, this remainder
contributes $O(\epsilon)\dd V^2$ to the metric and therefore
vanishes in the limit. The same suppression holds for any
fixed number of derivatives with respect to the rescaled
coordinates. Since the cross term and angular sector are
fully retained through second radial order, the second
order expansion is sufficient to recover the limiting
throat geometry.

The relation to the original string Carroll fields can also be followed in the coframe. After the near extremal scaling and time rescaling the coframe~\eqref{eq:rn-frame-exact} has $T^+=\dd V/\epsilon$ and a finite $T^-$. The physical null coframe is $(\sqrt\epsilon\,T^+,\sqrt\epsilon\,T^-)$. For each $\epsilon>0$, a Lorentz boost with factors $\sqrt\epsilon$ and $1/\sqrt\epsilon$ gives, with $q_\epsilon(R)=r_0+\epsilon(R+a/2)$,
\begin{equation}
 \begin{split}
 \vartheta^+&=\epsilon T^+=\dd V,\\
 \vartheta^-&=T^-=\dd R-\frac{R(R+a)}{2q_\epsilon(R)^2}\dd V
       \longrightarrow\dd R-\frac{F_a(R)}2\dd V,\\
 E^i&=q_\epsilon(R)\bar e^i\longrightarrow r_0\bar e^i.
 \end{split}
 \label{eq:finite-throat-frame}
\end{equation}
The boost preserves the metric since its two factors have product one. The limiting coframe is invertible, including at $R=0$, and its bilinears reproduce~\eqref{eq:throat}. The subleading longitudinal field identified in~\eqref{eq:rn-frame-coefficients} thus supplies the quadratic term in a regular Lorentzian throat frame.

This construction relates the string Carroll description
of the non extremal geometry to the curved Lorentzian throat.
Under the near extremal scaling and time rescaling, the
longitudinal sector, subleading at fixed non extremality,
contributes at leading order in the limiting metric.
The surviving second order temporal contribution supplies
its $\mathrm{AdS}_2$ curvature. The metric and the boosted
coframe have regular limits, although the boost becomes
singular as $\epsilon\to0$ and lies outside the regular
transformations considered in
appendix~\ref{app:sc-symmetries}. The regularity of the
limiting geometry therefore does not imply separate
regular limits for the original string Carroll coframes.

\section{The Maxwell field and the complete limiting solution}\label{sec:maxwell}

\subsection{The Maxwell potential: gauge choice, radial expansion and throat limit}

We take the electrically charged RN solution with a fixed sign of $Q$. Our normalization of the Einstein--Maxwell action and field strength is
\begin{equation}
 I=\frac{1}{16\pi G}\int\dd^4x\sqrt{-g}
 \left(\mathcal R-\cF_{\mu\nu}\cF^{\mu\nu}\right),\qquad
 \cF=\dd\mathcal{A}.
 \label{eq:em-action}
\end{equation}\\
A standard choice for the RN potential in ingoing coordinates is
$\mathcal{A}=-Q\,\mathrm{d}v/r$~\cite{Hubeny:1998ga}.
However,this potential does not admit a finite limit under the correlated scaling. After the time rescaling, its nonzero horizon value produces the constant term $-Q\,\dd V/(\epsilon r_+)$, which diverges as $\epsilon\to0$. This term can be removed by a gauge transformation without changing the electromagnetic field strength. We therefore shift the potential by $\dd(Qv/r_+)$ for each member of the RN family and use
\begin{equation}
 \begin{aligned}
 \mathcal{A}
 &=Q\left(\frac{1}{r_+}-\frac{1}{r_++x}\right)\dd v
   =\frac{Qx}{r_+(r_++x)}\,\dd v,
 \\[4pt]
 \cF
 &=\frac{Q}{(r_++x)^2}\,\dd x\wedge\dd v.
 \end{aligned}
 \label{eq:potential}
\end{equation}
This potential is regular and vanishes at the outer horizon, and it admits a finite limit under the correlated scaling.

At fixed $r_+$ and $\delta$, the potential has the radial expansion
\begin{equation}
 \mathcal{A}
 =\left(
 \epsilon\frac{QR}{r_+^2}
 -\epsilon^2\frac{QR^2}{r_+^3}
 +O(\epsilon^3)
 \right)\dd v.
 \label{eq:potential-expansion}
\end{equation}
In the notation of the background expansion in Appendix~\ref{app:sc-symmetries}, this corresponds to $\mathcal{A}_0=0$, $\mathcal{A}_1=QR\,\dd v/r_+^2$ and $\mathcal{A}_2=-QR^2\,\dd v/r_+^3$. These are coefficients of the Maxwell potential in the chosen  gauge.

Applying~\eqref{eq:scaling} to the radial expansion~\eqref{eq:potential-expansion}, with $r_+\to r_0$ and $Q\to Q_0$, gives the limiting potential and field strength
\begin{equation}
 \mathcal{A}_a=\frac{Q_0R}{r_0^2}\,\dd V,
 \qquad
 \cF_a=\frac{Q_0}{r_0^2}\,\dd R\wedge\dd V,
 \label{eq:limiting-maxwell}
\end{equation}
where $Q_0^2=r_0^2$. These expressions are independent of $a$ in the ingoing coordinates. The electric field is present both for the finite temperature patch and for the extremal member.\\
Only the first order term in \eqref{eq:potential-expansion} survives the limit~\eqref{eq:scaling}, giving the limiting potential and field strength in~\eqref{eq:limiting-maxwell}. As shown in Section~\ref{Einstein--Maxwell consistency}, these fields, together with the throat metric~\eqref{eq:throat}, form a complete local Einstein--Maxwell solution for every $a\geq0$. The metric and the potential thus require different original radial orders. An order assigned to a potential also depends on its gauge; the field strength and the limiting equations give the gauge invariant check. 

\subsection{Einstein--Maxwell consistency of the throat limit}
\label{Einstein--Maxwell consistency}
The equations following from~\eqref{eq:em-action} are
\begin{equation}
 \nabla_\mu\cF^{\mu\nu}=0,\qquad \dd\cF=0,\qquad
 \mathcal R_{\mu\nu}=2\left(\cF_{\mu\lambda}\cF_\nu{}^\lambda
 -\frac14g_{\mu\nu}\cF_{\alpha\beta}\cF^{\alpha\beta}\right).
 \label{eq:em-equations}
\end{equation}
For~\eqref{eq:throat} and~\eqref{eq:limiting-maxwell},
\begin{equation}
 \cF_{\mu\nu}\cF^{\mu\nu}=-\frac{2Q_0^2}{r_0^4}=-\frac{2}{r_0^2}.
 \label{eq:maxwell-invariant}
\end{equation}
The field strength is proportional to the volume form of the longitudinal factor, so it is closed and obeys the Maxwell equation. Writing $g_{\parallel}$ for the longitudinal metric and $g_{S^2}$ for the sphere metric, their Ricci tensors are
\begin{equation}
 \operatorname{Ric}(g_{\parallel})=-\frac{1}{r_0^2}g_{\parallel},\qquad
 \operatorname{Ric}(g_{S^2})=\frac{1}{r_0^2}g_{S^2},\qquad \mathcal R=0.
 \label{eq:product-ricci}
\end{equation}
Here $\mathcal R$ denotes the Ricci scalar of the full four dimensional throat geometry.
The field strength~\eqref{eq:limiting-maxwell} has only longitudinal components, giving $\cF_{\mu\lambda}\cF_\nu{}^\lambda=-(Q_0^2/r_0^4)(g_{\parallel})_{\mu\nu}$. Using this relation and~\eqref{eq:maxwell-invariant}, the electromagnetic contribution to Einstein's equation becomes
\[
2\left(
\cF_{\mu\lambda}\cF_\nu{}^\lambda
-\frac14g_{\mu\nu}\cF_{\alpha\beta}\cF^{\alpha\beta}
\right)
=
-\frac{1}{r_0^2}(g_{\parallel})_{\mu\nu}
+\frac{1}{r_0^2}(g_{S^2})_{\mu\nu}.
\]
This reproduces the longitudinal and sphere Ricci tensors in~\eqref{eq:product-ricci}, while the mixed components vanish on both sides. The limiting pair is therefore a complete local Einstein--Maxwell solution for every $a\geq0$.

The first order limiting metric~\eqref{eq:first-limit} has a flat longitudinal factor and the transverse two sphere of radius $r_0$. Its four dimensional scalar curvature is $2/r_0^2$, whereas source free Einstein--Maxwell theory requires $\mathcal R=0$. Even though the scaled electromagnetic field has a regular limit, it cannot make this metric an Einstein--Maxwell solution. The second order temporal term supplies the negative curvature required to balance the sphere curvature and the electric stress. This is an on shell consequence of the missing term in addition to the longitudinal curvature diagnostic.

\section{Analysis in static coordinates}
\label{sec:static}

We now repeat the near horizon analysis in static coordinates.
The temporal component again requires the second order
contribution, while the radial component involves terms from
all orders of its non extremal expansion.

\subsection{The string Carroll metric and its limit}

The static time coordinate is related to the ingoing coordinate
by $\dd v=\dd t+\dd x/f(x)$. With $x=\epsilon R$, the RN metric
takes the form
\begin{equation}
 \begin{split}
 \dd s^2={}&
 -\frac{\epsilon R(\delta+\epsilon R)}
 {(r_++\epsilon R)^2}\dd t^2\\
 &+\epsilon\frac{(r_++\epsilon R)^2}
 {R(\delta+\epsilon R)}\dd R^2
 +(r_++\epsilon R)^2\dd\Omega_2^2.
 \end{split}
 \label{eq:static-seed}
\end{equation}
We work in the exterior region $R>0$. Expanding at fixed
$r_+$ and $\delta>0$, the metric retained through first
radial order is\footnote{The label $\mathrm{st}$ denotes the static coordinate chart.}
\begin{equation}
 \begin{split}
 \dd s_{\mathrm{st},[1],\epsilon}^2
 ={}&r_+^2\dd\Omega_2^2\\
 &+\epsilon\left[
 -\frac{\delta R}{r_+^2}\dd t^2
 +\frac{r_+^2}{\delta R}\dd R^2
 +2r_+R\dd\Omega_2^2
 \right].
 \end{split}
 \label{eq:static-first}
\end{equation}
This is the string Carroll metric in the static chart.\\
We now apply the near extremal scaling
$\delta=\epsilon a$ and $r_\pm=r_0\pm\epsilon a/2$,
with the static time rescaled as $t=T/\epsilon$.
Taking $\epsilon\to0$ at fixed $r_0>0$, $a>0$,
and fixed $T,R,\theta,\phi$ gives
\begin{equation}
 \dd s_{\mathrm{st},[1],a}^2
 =-\frac{aR}{r_0^2}\dd T^2
 +\frac{r_0^2}{aR}\dd R^2
 +r_0^2\dd\Omega_2^2.
 \label{eq:static-first-limit}
\end{equation}
The angular correction vanishes in this limit.
The longitudinal metric has
$F_{[1],a}(R)=aR/r_0^2$ and hence
$\mathcal R^{(2)}=-F_{[1],a}''(R)=0$.
Thus the first order static metric gives the same
Rindler geometry found in the ingoing analysis and
does not recover the $\mathrm{AdS}_2$ throat.

\subsection{The second order near horizon expansion}

At fixed non extremal parameters, retaining the next
radial order gives
\begin{equation}
 \begin{split}
 \dd s_{\mathrm{st},[2],\epsilon}^2
 ={}&\dd s_{\mathrm{st},[1],\epsilon}^2\\
 &+\epsilon^2\biggl[
 -R^2\left(\frac{1}{r_+^2}
          -\frac{2\delta}{r_+^3}\right)\dd t^2
 +\left(\frac{2r_+}{\delta}
        -\frac{r_+^2}{\delta^2}\right)\dd R^2
 +R^2\dd\Omega_2^2
 \biggr].
 \end{split}
 \label{eq:static-second}
\end{equation}
The temporal and angular contributions agree with those
in the ingoing expansion. The radial contribution now
contains inverse powers of the horizon separation
$\delta$.

Under the same near extremal scaling and time rescaling,
the second order temporal term contributes
$-R^2\dd T^2/r_0^2$, while the second order radial term
contributes $-r_0^2\dd R^2/a^2$. The angular contribution
vanishes. The resulting expression is therefore
\begin{equation}
 \begin{split}
 \dd s_{\mathrm{st},[2],a}^2
 ={}&-\frac{R(R+a)}{r_0^2}\dd T^2\\
 &+\left(\frac{r_0^2}{aR}
         -\frac{r_0^2}{a^2}\right)\dd R^2
 +r_0^2\dd\Omega_2^2.
 \end{split}
 \label{eq:static-second-limit}
\end{equation}\\
The temporal component now agrees with that of the
throat obtained in~\eqref{eq:throat}. To compare the
radial components, write the latter metric in static
coordinates using
$\dd V=\dd T+\dd R/F_a(R)$. Its radial coefficient is
\begin{equation}
 G(R,a)=\frac{1}{F_a(R)}
       =\frac{r_0^2}{R(R+a)}.
 \label{eq:static-exact}
\end{equation}
The radial coefficient
$r_0^2/(aR)-r_0^2/a^2$ in
\eqref{eq:static-second-limit} contains only the first
two terms of the expansion of $G(R,a)$ in $R/a$.
Consequently, the second order expansion recovers the temporal
component but does not suffice for the static radial
component.

\subsection{Radial contributions from all orders}

The origin of the remaining contributions can be seen
directly in the denominator of the radial term
in~\eqref{eq:static-seed}. At fixed $\delta>0$,
\begin{equation}
 \frac{1}{\delta+\epsilon R}
 =\frac{1}{\delta}
  \sum_{n=0}^{\infty}
  \left(-\frac{\epsilon R}{\delta}\right)^n,
 \qquad |\epsilon R|<\delta.
 \label{eq:geometric}
\end{equation}
The radial coefficient therefore has the expansion
\begin{equation}
 g_{RR}^{\mathrm{st}}
 =
 \frac{\epsilon}{R\delta}
 \left(r_+^2+2\epsilon r_+R+\epsilon^2R^2\right)
 \sum_{n=0}^{\infty}
 \left(-\frac{\epsilon R}{\delta}\right)^n.
 \label{eq:static-radial-series}
\end{equation}\\
Consider first the contributions proportional to $r_+^2$.
The term of original radial order $\epsilon^{n+1}$
satisfies
\begin{equation}
 (-1)^n\epsilon^{n+1}
 \frac{r_+^2R^{n-1}}{\delta^{n+1}}
 \longrightarrow
 \frac{r_0^2}{aR}
 \left(-\frac{R}{a}\right)^n.
 \label{eq:static-surviving-orders}
\end{equation}
Thus a contribution survives from every radial order.
The terms arising from $2\epsilon r_+R$ and
$\epsilon^2R^2$ in the numerator retain an additional
factor of $\epsilon$ and $\epsilon^2$, respectively,
and vanish in the limit.

The expansion parameter makes the change in behaviour
explicit. At fixed non extremality, $\epsilon R/\delta$
vanishes as the horizon is approached. Under
$\delta=\epsilon a$, however, it becomes $R/a$, which
does not decrease with $\epsilon$. Higher radial orders
are therefore not suppressed by the limiting procedure.

For $0<R<a$, the geometric series converges, summing the surviving contributions gives
\begin{equation}
 \lim_{\epsilon\to0}
 g_{RR}^{\mathrm{st}}
 =
 \frac{r_0^2}{aR}
 \sum_{n=0}^{\infty}
 \left(-\frac{R}{a}\right)^n
 =
 \frac{r_0^2}{R(R+a)}.
 \label{eq:static-resummation}
\end{equation}
The resulting expression agrees with the radial coefficient of the near extremal RN throat metric in static coordinates, given in ~\eqref{eq:static-exact} throughout $R>0$.
Outside $0<R<a$, this rational expression supplies the continuation; the geometric series itself does not converge.\\
Combining the radial result with the temporal and angular
limits gives
\begin{equation}
 \dd s_{\mathrm{st},a}^2
 =
 -\frac{R(R+a)}{r_0^2}\dd T^2
 +\frac{r_0^2}{R(R+a)}\dd R^2
 +r_0^2\dd\Omega_2^2.
 \label{eq:static-throat}
\end{equation}
This is the static form of~\eqref{eq:throat}.\\
Setting $a=0$ in~\eqref{eq:static-throat} gives the
extremal throat in static coordinates.

\section{Discussion and outlook}\label{sec:discussion}

Motivated by the question of how the string Carroll geometry of non extremal horizons naturally evolves to a Lorentzian geometry as extremality is approached~\cite{Bagchi:2024rje,Bagchi:2026qpi}, we have established this connection for RN. We identified the quadratic temporal contribution that is suppressed in the non extremal description but is required to recover the \(\mathrm{AdS}_2\) curvature. Approaching the outer horizon and extremality simultaneously, together with the corresponding rescaling of Killing time, brings this contribution into the leading longitudinal metric alongside the Rindler contribution, recovering the Lorentzian \(\mathrm{AdS}_2\times S^2\) throat.

An immediate application concerns probe dynamics. Existing string Carroll analyses derive classical string actions, constraints and solutions near non extremal horizons. For Schwarzschild, magnetic strings collapse to a point on the leading transverse sphere and exhibit longitudinal Rindler motion and angular fluctuations at subleading order in the Polyakov expansion, while electric strings can retain angular winding~\cite{Bagchi:2023cfp,Bagchi:2024rje}. These sectors have also been investigated in BTZ backgrounds~\cite{Banerjee:2025bkg}. More recently, chiral strings have been studied, with angular fluctuations of one chirality coupled to longitudinal motion~\cite{Kolekar:2026Chiral}. Particle and scalar probes provide further examples in which actions and equations have been organized directly in the non extremal string Carroll expansion~\cite{Bagchi:2026qpi}.

The present construction provides the background for investigating how these descriptions change towards extremality. Applying the same correlated scaling to the first order RN string Carroll expansion and to its extension used in the present work gives different limiting geometries: the former has a flat longitudinal sector, while the latter retains the curved throat. For probes with regular limits in the rescaled coordinates, this difference enters the leading dynamics. In particular, for a finite tension string with a smooth embedding and a regular limiting worldsheet metric, the subleading longitudinal terms and the further suppressed temporal contribution in the non extremal string Carroll background enter the leading Polyakov action, embedding equations and Virasoro constraints alongside the leading angular geometry after the correlated limit. 
Extending the existing string Carroll descriptions of probe dynamics to the throat regime thus requires a consistent treatment of the embeddings, momenta and worldsheet auxiliary fields, together with the chosen tension scaling. This would determine how the magnetic, electric and chiral sectors reorganize, and how their constraints and fluctuation equations change. Scalar propagation offers another concrete comparison: angular derivatives and the mass term enter beyond the leading longitudinal kinetic term in the non extremal expansion~\cite{Bagchi:2026qpi}. Following these contributions through the correlated limit would establish their role in the leading throat theory. The same analysis could determine the first corrections to the limiting dynamics and the additional background coefficients they require.

Beyond probe dynamics, the RN reconstruction developed here and its extension to smooth stationary families admitting regular extremal limits provide a basis for investigating the approach to extremality in other applications of string Carroll geometry~\cite{Shinde:2026GeneralThroats}. The formulation of string Carroll gravity for non extremal rotating black holes offers a concrete setting in which to determine how the gravitational action and field equations must be reorganized under the correlated scaling~\cite{Bhattacharya:2026UnifiedGravity}. A complementary application is spin hydrodynamics, where stationary transport and conserved matter fluxes have been calculated near non extremal spherical horizons, including RN~\cite{Lobos:2026SpinTransport}. Extending these calculations towards extremality requires following the dynamical fields together with the background geometry. The reconstruction supplies a geometric starting point for determining whether the orders retained in the non extremal description remain sufficient for the physical quantity under consideration, and which additional contributions must be included.

\acknowledgments
We thank the people of India for their continuous support for science. We thank IIT(ISM) Dhanbad and the Department of Physics for providing us with the academic infrastructure necessary for our research work.

\appendix
\section{Longitudinal curvature}\label{app:curvature}

For\footnote{The subscript $2$ in $\dd s_2^2$ refers to the dimension
of the longitudinal geometry.} $\dd s_2^2=-F(R)\dd V^2+2\dd V\dd R$, the nonzero inverse components are $g^{VR}=g^{RV}=1$ and $g^{RR}=F$. The nonzero connection coefficients are
\begin{equation}
 \Gamma^V_{VV}=\frac{F'}2,\qquad
 \Gamma^R_{VV}=\frac{FF'}2,\qquad
 \Gamma^R_{VR}=\Gamma^R_{RV}=-\frac{F'}2.
\end{equation}
They give $\mathcal R_{VV}=FF''/2$, $\mathcal R_{VR}=-F''/2$ and $\mathcal R_{RR}=0$, hence
\begin{equation}
 \mathcal R^{(2)}=-F''.
\end{equation}
This derivation holds in the regular ingoing chart, including at a zero of $F$.

\section{Transformations of the expanded background fields}
\label{app:sc-symmetries}

The coframe decomposition used in the main text admits local changes
of frame that leave the metric unchanged. In this appendix, we derive
the corresponding transformations of the expansion coefficients,
together with their diffeomorphism and Maxwell gauge transformations.
The coframe laws are obtained by expanding the local Lorentz and diffeomorphism transformations.
Throughout this appendix, we use the identification $\epsilon=c^2$ introduced in the main text.\\
We retain the coframe and Maxwell potential through order $\epsilon^2$:
\begin{equation}
\begin{aligned}
T_\epsilon
&=\tau+\epsilon m+\epsilon^2n+O(\epsilon^3),\\
E_\epsilon
&=e+\epsilon e_{(1)}+\epsilon^2e_{(2)}+O(\epsilon^3),\\
\mathcal A_\epsilon
&=\mathcal A_0+\epsilon\mathcal A_1+\epsilon^2\mathcal A_2
+O(\epsilon^3),
\qquad
\mathcal F_\epsilon=\mathrm d\mathcal A_\epsilon.
\end{aligned}
\label{eq:app-field-jets}
\end{equation}
Here and below, longitudinal frame indices $A,B=0,1$ and transverse
frame indices $i,j=2,3$ are suppressed where no confusion arises.
The coefficient $n^A$ completes the longitudinal coframe
expansion~\eqref{eq:sc-frame-fields} at this order. Owing to the overall
factor of $\epsilon$ in the longitudinal metric, it first contributes
to $g_\epsilon$ at order $\epsilon^3$ and does not enter $k_2$.

Let $\mathrm Q_\epsilon$ denote the coframe matrix, whose rows carry
frame indices and whose columns carry spacetime indices. Its entries
are $T^A_{\epsilon\mu}$ and $E^i_{\epsilon\mu}$, and we assume that
$\mathrm Q_0$ is invertible. The metric can then be written as
\[
g_\epsilon=\mathrm Q_\epsilon^T\mathrm D_\epsilon\mathrm Q_\epsilon,
\qquad
\mathrm D_\epsilon=\operatorname{diag}(\epsilon\eta,I_2),
\qquad
\eta=\operatorname{diag}(-1,1),
\]
where $I_2$ is the identity matrix in the transverse sector.

An infinitesimal frame transformation
$\delta_{\mathrm{frame}}\mathrm Q_\epsilon
=\mathrm M_\epsilon\mathrm Q_\epsilon$
induces
\[
\delta_{\mathrm{frame}}g_\epsilon
=
\mathrm Q_\epsilon^T
\left(
\mathrm M_\epsilon^T\mathrm D_\epsilon
+\mathrm D_\epsilon\mathrm M_\epsilon
\right)
\mathrm Q_\epsilon.
\]
Since the coframe matrix is invertible, preservation of the metric requires
\[
\mathrm M_\epsilon^T\mathrm D_\epsilon
+\mathrm D_\epsilon\mathrm M_\epsilon=0.
\]
For $\epsilon>0$, solving this condition in longitudinal and transverse
blocks gives
\begin{equation}
\begin{aligned}
\mathrm M_\epsilon
&=
\begin{pmatrix}
L_\epsilon&\Lambda_\epsilon\\
\epsilon C_\epsilon&S_\epsilon
\end{pmatrix},
\qquad
C_\epsilon=-\Lambda_\epsilon^T\eta,
\\[3pt]
L_\epsilon^T\eta+\eta L_\epsilon&=0,
\qquad
S_\epsilon^T+S_\epsilon=0.
\end{aligned}
\label{eq:app-frame-generator}
\end{equation}
These conditions identify $L_\epsilon$ as the longitudinal Lorentz
generator and $S_\epsilon$ as the transverse rotation generator.
The parameter $\Lambda_\epsilon$ describes the mixing between the two
sectors, with the other mixed block fixed by metric preservation.
The relative factor of $\epsilon$ between the mixed blocks follows
from the relative weighting of the longitudinal and transverse sectors
in the metric.\\
Including a diffeomorphism generated by $\xi_\epsilon$ and a Maxwell
gauge transformation with scalar parameter $\alpha_\epsilon$, the fields transform as
\begin{equation}
\begin{aligned}
\delta\mathrm Q_\epsilon
&=\operatorname{Lie}_{\xi_\epsilon}\mathrm Q_\epsilon
+\mathrm M_\epsilon\mathrm Q_\epsilon,\\
\delta\mathcal A_\epsilon
&=\operatorname{Lie}_{\xi_\epsilon}\mathcal A_\epsilon
+\mathrm d\alpha_\epsilon.
\end{aligned}
\label{eq:app-exact-transformations}
\end{equation}
The Lie derivative acts on spacetime tensor indices, whereas
$\mathrm M_\epsilon$ acts on frame indices. Since $\mathcal A_\epsilon$
is a spacetime one form, it is unchanged by the internal frame
transformation. Its field strength is also unchanged by the Maxwell
gauge transformation, as $\mathrm d^2\alpha_\epsilon=0$.

We take the transformation parameters to be independent of the fields,
while allowing arbitrary dependence on the spacetime coordinates.
Their expansions contain only nonnegative integer powers of
$\epsilon$. Thus, for example,
\[
\Lambda_\epsilon=\sum_{p\geq0}\epsilon^p\Lambda_p,
\]
with the same convention for the remaining parameters and for
$\mathrm M_\epsilon$. The dual frame matrix is expanded as
\[
\mathrm P_\epsilon=\mathrm Q_\epsilon^{-1}
=\sum_{p\geq0}\epsilon^p\mathrm P_p.
\]
Its transformation follows by varying
$\mathrm Q_\epsilon\mathrm P_\epsilon=I_4$, which gives
\[
\delta\mathrm P_\epsilon
=\operatorname{Lie}_{\xi_\epsilon}\mathrm P_\epsilon
-\mathrm P_\epsilon\mathrm M_\epsilon.
\]\\
Substituting these expansions
into~\eqref{eq:app-exact-transformations} and collecting equal powers
of $\epsilon$ gives the coefficient transformations. With
\[
(T_0,T_1,T_2)=(\tau,m,n),
\qquad
(E_0,E_1,E_2)=(e,e_{(1)},e_{(2)}),
\]
they take the following form for $p=0,1,2$:
\begin{equation}
\begin{aligned}
\delta T_p
&=\sum_{q=0}^{p}
\left(
\operatorname{Lie}_{\xi_q}T_{p-q}
+L_qT_{p-q}
+\Lambda_qE_{p-q}
\right),\\[3pt]
\delta E_p
&=\sum_{q=0}^{p}
\left(
\operatorname{Lie}_{\xi_q}E_{p-q}
+S_qE_{p-q}
\right)
+\sum_{q=0}^{p-1}C_qT_{p-1-q},\\[3pt]
\delta\mathcal A_p
&=\sum_{q=0}^{p}
\operatorname{Lie}_{\xi_q}\mathcal A_{p-q}
+\mathrm d\alpha_p,\\[3pt]
\delta\mathcal F_p
&=\sum_{q=0}^{p}
\operatorname{Lie}_{\xi_q}\mathcal F_{p-q},\\[3pt]
\delta\mathrm P_p
&=\sum_{q=0}^{p}
\left(
\operatorname{Lie}_{\xi_q}\mathrm P_{p-q}
-\mathrm P_{p-q}\mathrm M_q
\right).
\end{aligned}
\label{eq:app-coefficient-transformations}
\end{equation}
Here $\mathcal F_p=\mathrm d\mathcal A_p$.
In~\eqref{eq:app-frame-generator}, $C_\epsilon$ enters the frame
generator as $\epsilon C_\epsilon$, so its coefficient in
$\mathrm M_p$ is $C_{p-1}$ for $p\geq1$.
The final sum in $\delta E_p$ arises from expanding
$\epsilon C_\epsilon T_\epsilon$; the explicit factor of $\epsilon$
accounts for the upper limit $p-1$ and the absence of this sum at $p=0$.\\
The induced transformations of the metric
coefficients~\eqref{eq:sc-k2} follow directly
from~\eqref{eq:app-frame-generator}. The internal contributions cancel,
leaving
\[
\delta g_\epsilon
=\operatorname{Lie}_{\xi_\epsilon}g_\epsilon.
\]
Expanding $g_\epsilon=h+\epsilon k_1+\epsilon^2k_2+O(\epsilon^3)$,
we obtain
\begin{equation}
\begin{aligned}
\delta h
&=\operatorname{Lie}_{\xi_0}h,\\
\delta k_1
&=\operatorname{Lie}_{\xi_0}k_1
+\operatorname{Lie}_{\xi_1}h,\\
\delta k_2
&=\operatorname{Lie}_{\xi_0}k_2
+\operatorname{Lie}_{\xi_1}k_1
+\operatorname{Lie}_{\xi_2}h.
\end{aligned}
\label{eq:app-metric-transformations}
\end{equation}
Thus each metric coefficient is invariant under internal
frame transformations, although its separate longitudinal and
transverse bilinears need not be. In particular, the expression
$2\tau^+\odot m^-$ for the temporal part of $k_2$ refers to the RN
coframe~\eqref{eq:rn-frame-coefficients} chosen in the main text.
Equation~\eqref{eq:app-metric-transformations} also shows how
diffeomorphisms with $\epsilon$ dependent parameters mix the metric
coefficients at different orders.

The coefficient transformations inherit their algebra from the
unexpanded laws. To see this, suppress the $\epsilon$ dependence and
introduce the differential operator
\[
\mathcal O=\operatorname{Lie}_\xi+\mathrm M
\]
acting on the coframe. For two parameter sets, labelled $[1]$ and
$[2]$, its operator commutator is
\begin{equation}
\begin{aligned}
[\mathcal O^{[1]},\mathcal O^{[2]}]
&=\operatorname{Lie}_{[\xi^{[1]},\xi^{[2]}]}
+\mathrm M^{[12]},\\
\mathrm M^{[12]}
&=\operatorname{Lie}_{\xi^{[1]}}\mathrm M^{[2]}
-\operatorname{Lie}_{\xi^{[2]}}\mathrm M^{[1]}
+[\mathrm M^{[1]},\mathrm M^{[2]}].
\end{aligned}
\label{eq:app-closure}
\end{equation}
Here $[\xi^{[1]},\xi^{[2]}]$ is the vector field Lie bracket, and
$\operatorname{Lie}_\xi\mathrm M=\xi^\mu\partial_\mu\mathrm M$
acts on the scalar matrix entries. The resulting matrix
$\mathrm M^{[12]}$ again satisfies the conditions
in~\eqref{eq:app-frame-generator}. In the same operator convention,
the Maxwell gauge parameter is
\[
\alpha^{[12]}
=
\operatorname{Lie}_{\xi^{[1]}}\alpha^{[2]}
-\operatorname{Lie}_{\xi^{[2]}}\alpha^{[1]}.
\]
Since the transformations contain neither negative powers of
$\epsilon$ nor derivatives with respect to it, terms beyond
$\epsilon^2$ cannot enter the retained coefficient laws. These laws
therefore close through the stated order and inherit the Jacobi
identity of the unexpanded algebra.\\
Finally, the order required in the coframe depends on the quantity
being constructed. In
\[
g_\epsilon^{-1}
=\mathrm P_\epsilon\mathrm D_\epsilon^{-1}\mathrm P_\epsilon^T,
\]
the longitudinal block of $\mathrm D_\epsilon^{-1}$ carries a factor
of $\epsilon^{-1}$. Coframe coefficients through $\epsilon^2$
therefore determine the inverse metric through the term proportional
to $\epsilon$; its $\epsilon^2$ coefficient generally requires the
longitudinal dual frame through $\epsilon^3$. The coefficients needed
for a probe action or its constraints must consequently be determined
by expanding those quantities separately.

\bibliographystyle{JHEP}
\bibliography{ref}
\end{document}